\documentclass[12pt,a4paper]{article}
\usepackage{amsmath,amssymb}
\usepackage{graphicx}
\begin{document}
\newcounter{Series}
\setcounter{Series}{1}
\title{
Revival of Single-Particle Transport Theory 
for the Normal State of High-$T_c$ Superconductors: \\
\Roman{Series}. 
Relaxation-Time Approximation }

\author{
O. Narikiyo
\footnote{
Department of Physics, 
Kyushu University, 
Fukuoka 812-8581, 
Japan}
}

\date{
(Jan. 17, 2014)
}

\maketitle
\begin{abstract}
How the fluctuation-exchange (FLEX) approximation 
and the Fermi-liquid theory fail to explain 
the anomalous behavior of the Hall coefficient 
in the normal state of high-$T_c$ superconductors 
is clarified. 
\end{abstract}
\vskip 30pt

A revival of a single-particle transport theory 
for the normal state of high-Tc superconductors 
is now in progress \cite{Anderson,KM,KHM}. 

In this Short Note 
I try to support the progress using a simple formula 
for the DC Hall conductivity. 
The formula is so simple 
that it can be implemented on your PC within ten minutes. 

The target data is the non-monotonic temperature dependence of 
the DC Hall conductivity 
measured in the normal state of PCCO superconductors \cite{Drew}. 
Although it was interpreted \cite{Drew} 
by the effect of the vertex correction, 
I shall show that it is explained qualitatively 
within the relaxation-time approximation. 

In the next Short Note 
I shall discuss why the relaxation-time approximation is enough 
for the qualitative understanding 
and the vertex correction is irrelevant. 

\newpage 

We start from the full Green function for electrons 
\begin{equation}
G^R({\bf p},\epsilon) 
= {1 \over \epsilon - E({\bf p},\epsilon) + i/2\tau({\bf p},\epsilon)}, 
\ \ \ \ \ \ 
G^A({\bf p},\epsilon) 
= {1 \over \epsilon - E({\bf p},\epsilon) - i/2\tau({\bf p},\epsilon)}, 
\label{G} 
\end{equation}
where the self-energy is renormalized into 
the dispersion $E({\bf p},\epsilon)$ and 
the life-time $\tau({\bf p},\epsilon)$. 

Our numerical calculation is done for 2D square lattice 
with the lattice constant $a \equiv 1$. 
By the symmetry we only need the information 
for the quarter of the Brillouin zone: 
$0 < p_x < \pi$ and $0 < p_y < \pi$. 
In the following 
the summation over ${\bf p}$ is restricted within this quarter. 

We consider the conductivity within the relaxation-time approximation. 
In the next Short Note 
we shall discuss  that the vertex correction does not change 
the qualitative result of the relaxation-time approximation 
in the case where the forward scattering is unimportant. 
Neglecting the vertex correction 
the DC conductivity $\sigma_{xx}$ in the absence of the magnetic field 
is given by \cite{FEW} 
\begin{equation}
\sigma_{xx} 
= {e^2 \over \pi} \cdot {4 \over R(T)}, 
\label{sigma_xx} 
\end{equation}
with 
\begin{equation}
{1 \over R(T)} 
\equiv \sum_{\bf p} v_x^2 G^R G^A 
= \sum_{\bf p} v_x^2 {1 \over E^2 + (1/2\tau)^2}, 
\label{R(T)} 
\end{equation}
where 
$G^R=G^R({\bf p},0)$, 
$G^A=G^A({\bf p},0)$, 
$E=E({\bf p},0)$, 
$\tau=\tau({\bf p},0)$ and 
$v_x=\partial E / \partial p_x$. 
Here we have assumed the Fermi degeneracy for simplicity. 

In the same manner 
the DC Hall conductivity $\sigma_{xy}$ 
proportional to the weak magnetic field $H$ 
is given by \cite{FEW} 
\begin{equation}
\sigma_{xy} / H 
= {|e|^3 \over \pi} \cdot 4 S(T), 
\label{sigma_xy} 
\end{equation}
with 
\begin{align}
S(T) 
& \equiv \sum_{\bf p} 
v_x \Big( v_x {\partial v_y \over \partial p_y} 
        - v_y {\partial v_x \over \partial p_y} \Big) 
G^R G^A { G^R - G^A \over 2 i } 
\nonumber \\ 
& = - \sum_{\bf p} 
v_x \Big( v_x {\partial v_y \over \partial p_y} 
        - v_y {\partial v_x \over \partial p_y} \Big) 
{1/2\tau \over \big[ E^2 + (1/2\tau)^2 \big]^2}, 
\label{S(T)} 
\end{align}
where $v_y=\partial E / \partial p_y$. 

\newpage 

Thus, if you have the data of the momentum dependence of 
the dispersion $E$ and the life-time $\tau$ at $\epsilon=0$, 
you can easily get the DC Hall conductivity from this formula. 
The implementation of the formula on your PC takes little effort. 
I employ the following model for $E$ and $\tau$. 

Here we substitute the dispersion obtained by the band calculation \cite{NKS} 
\begin{equation}
E = -2t(\cos p_x + \cos p_y) + 4t'\cos p_x \cdot \cos p_y 
    -2t''(\cos 2p_x + \cos 2p_y) - \mu, 
\label{E(p)} 
\end{equation}
for that for quasi-particles. 
For PCCO we adopt $t=-0.438$eV, $t'=0.156$eV and $t''=0.098$eV \cite{NKS} 
in the numerical calculations in this Short Note. 
By this substitution we miss some many-body correlations. 

The Fermi surface for this dispersion is shown in Fig.~1. 
The factor $f({\bf p}) \equiv 
 v_x ( v_x \partial v_y / \partial p_y 
     - v_y \partial v_x / \partial p_y ) $ 
appearing in $\sigma_{xy}$ is shown in Fig.~2. 

For the life-time we adopt the model 
\begin{equation}
{1 \over 2\tau} = 
w({\bf p})\cdot g_1 \cdot T 
+ \big[ 1 - w({\bf p}) \big]\cdot g_2 \cdot T^2, 
\label{tau} 
\end{equation}
similar to the multi-patch model \cite{PSK} 
where the Brillouin zone is divided into hot and cold patches. 
The life-time is determined 
by the coupling to the anti-ferromagnetic spin-fluctuation. 
In the cold patches 
the relevant spin-fluctuation spectrum is broad 
and its temperature-dependence is weak 
so that $1/\tau \sim T^2$. 
On the other hand, in the hot patches 
it is peaked around the nesting-vector 
and its integrated weight depends on the temperature 
so that $1/\tau \sim T$. 
For simplicity 
we did not implement the gradual change 
between hot and cold patches employed in \cite{PSK}. 
Namely $w({\bf p})$ is a step function in our case 
where $w({\bf p})=1$ 
if $({\bf p}-{\bf p}_1)^2 < r^2$ or $({\bf p}-{\bf p}_2)^2 < r^2$ 
with ${\bf p}_1 \equiv (\pi,0)$ and ${\bf p}_2 \equiv (0,\pi)$ 
and $w({\bf p})=0$ otherwise. 

We set $g_2 = 100/$eV in accordance with \cite{KHM}. 
Here the temperature $T$ is measured in eV. 
On the other hand, our choice, $g_1 = 0.1$, is smaller 
than \cite{KHM} and \cite{PSK}. 
Such a smallness is explained by the nested spin fluctuation \cite{NM}. 

\newpage 

Before performing the numerical calculation 
we can make a rosy prediction 
that the non-monotonic temperature dependence of $\sigma_{xy}$, 
which we want to derive, is obtained, 
if the temperature dependences of $1/\tau$ are different 
between the regions with positive $f({\bf p})$ 
and the regions with negative $f({\bf p})$ 
in the Brillouin zone. 
After the numerical calculation 
we confirm the prediction as shown in Fig.~3. 
The result for $\mu=0$ actually shows 
the non-monotonic temperature dependence 
qualitatively similar to that observed in the experiment \cite{Drew}. 
More direct comparison between the experiment \cite{Drew} 
and our numerical calculation can be done 
for $\sigma_{xy}/\sigma_{xx}$ shown in Fig.~4. 

In conclusion our simple formula 
based on the relaxation-time approximation for single particles 
qualitatively explain the non-monotonic behavior of the DC Hall conductivity 
in the normal state of PCCO superconductors, 
though our choice of the parameters 
should be reconsidered for quantitative description. 

This work was driven by the discussions 
with Professor Kazumasa Miyake. 

\vskip 30pt
\noindent
{\bf \Large Appendix}
\vskip 12pt

In the main text of this Short Note 
the anomalous behavior of the Hall coefficient 
in the normal state of high-$T_c$ superconductors 
is explained within the relaxation-time approximation\footnote{
This terminology might not be appropriate. 
Precisely speaking 
we have only neglected the current-vertex correction (CVC). }. 

In this Appendix 
how the fluctuation-exchange (FLEX) approximation 
and the Fermi-liquid theory fail to explain 
the anomalous behavior of the coefficient (ABC) is clarified. 

Let us start with the formulae \cite{FEW} 
for $\sigma_{xx}$ 
\begin{equation}
\sigma_{xx} 
= {4e^2 \over \pi} \sum_{\bf p} v_x^2 G^R G^A 
= {4e^2 \over \pi} \sum_{\bf p} v_x^2 {1 \over E^2 + (1/2\tau)^2}, 
\label{sigma_xxA} 
\end{equation}
and for $\sigma_{xy}$ 
\begin{align}
\sigma_{xy} / H 
& = {4|e|^3 \over \pi} \sum_{\bf p} 
v_x \Big( v_x {\partial v_y \over \partial p_y} 
        - v_y {\partial v_x \over \partial p_y} \Big) 
G^R G^A { G^R - G^A \over 2 i } 
\nonumber \\ 
& = - {4|e|^3 \over \pi} \sum_{\bf p} 
v_x \Big( v_x {\partial v_y \over \partial p_y} 
        - v_y {\partial v_x \over \partial p_y} \Big) 
{1/2\tau \over \big[ E^2 + (1/2\tau)^2 \big]^2}. 
\label{sigma_xyA} 
\end{align}
Here the propagators $G^R$ and $G^A$ 
are expressed in terms of the spectral function 
$\rho_{\bf p}(\varepsilon)$. 

As shown in \cite{FEW} 
these formulae lead to the constant Hall coefficient 
if the spectral function $\rho_{\bf p}(\varepsilon)$ is delta-function-like 
and isotropic\footnote{
Although some anisotropy leads to 
weak temperature dependence of the Hall coefficient 
as discussed in arXiv:cond-mat/0006028, 
it is too weak to explain the anomalous temperature dependence 
observed in experiments. } 
in momentum space. 
The Fermi-liquid theory for the Hall coefficient \cite{KY} also assumes 
the delta-function-like spectral function 
so that it also leads to almost temperature-independent Hall coefficient. 
Thus the Fermi-liquid theory fails to explain the ABC. 

The above-mentioned delta-function-like spectral function 
is justified only for weakly correlated systems. 
Our calculation in the main text 
is performed without such a delta-function assumption. 
We have to use a broad spectral function\footnote{
A schematic spectral function is shown in Fig.~5. } 
in a finite momentum space (the 1st Brillouin zone) 
in the case of strongly correlated system. 

Although our spectral function employed in the main text is a simple model, 
it captures the features of actual one observed by the ARPES experiments. 
Thus our result for the Hall angle in Fig.~4 
qualitatively explains the anomalous non-monotonic temperature dependence 
observed in experiments. 
Thus we can conclude 
that the ABC is explained without the CVC 
if we employ the correct spectral function. 

On the other hand, 
the Hall angle calculated 
without the CVC in the FLEX approximation, 
the inset of Fig.~2 in \cite{Drew}, 
is totally different from the one observed by experiments. 
The failure of this calculation does not mean the necessity of the CVC. 
It only means 
that the spectral function obtained by the FLEX approximation is incorrect. 
We have already discussed in arXiv:1301.5996 
that the FLEX approximation is not applicable 
to the system with the Fermi degeneracy. 

Consequently, the spectral function is the key to explain the ABC. 
The FLEX approximation fails to explain the ABC, 
since the correct spectral function is not obtained by this approximation. 
The Fermi-liquid theory fails to explain the ABC, 
since the delta-function assumption is not justified 
for strongly correlated systems at the room temperature. 

This Appendix arose from the two seminars 
held at the department of physics and the elementary-particle-theory group 
of Kyushu University. 
I thank to the organizers of these seminars. 



\newpage 

\begin{figure}[htbp]
\begin{center}
\includegraphics[width=11.6cm]{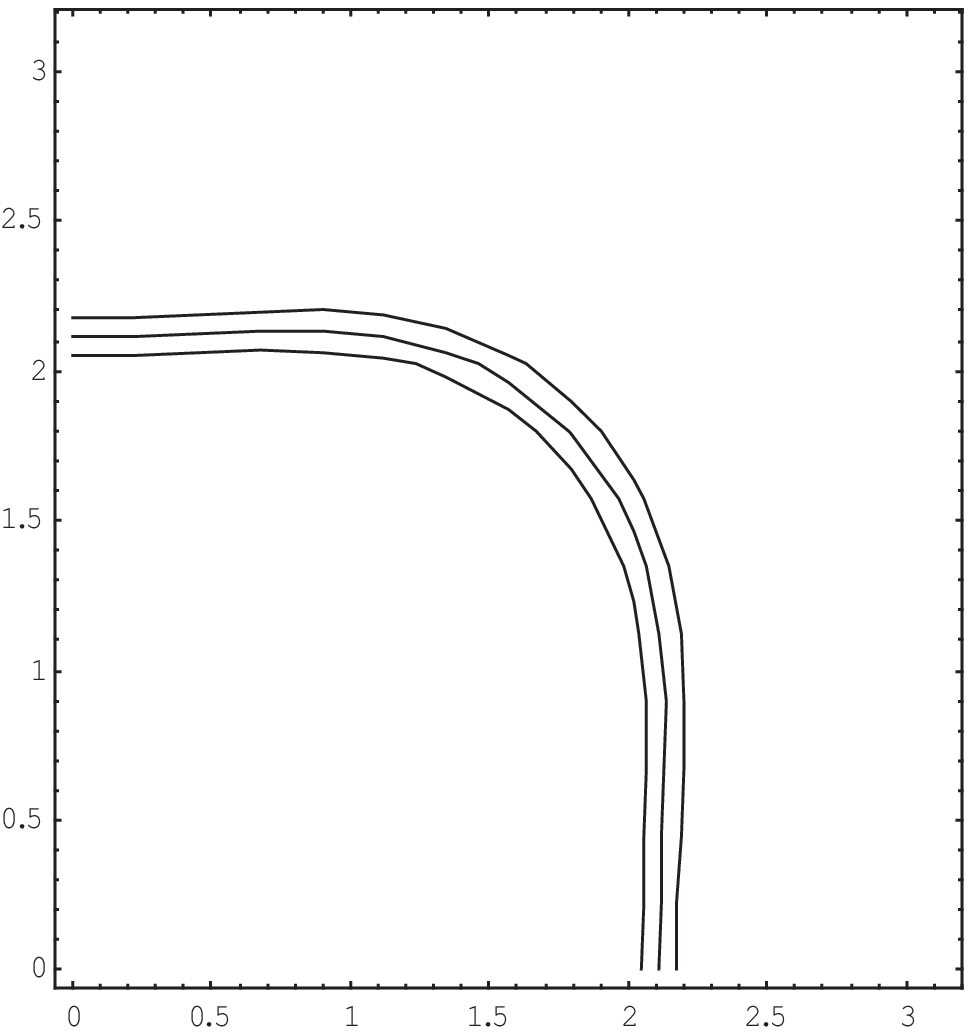}
\vskip 16mm
\caption{Fermi surfaces for $\mu=-0.1$, $0$, $0.1$ 
with $t=-0.438$, $t'=0.156$, $t''=0.098$ 
where all the energies are represented in eV. 
Only the quarter of the Brillouin zone 
($0 < p_x < \pi$ and $0 < p_y < \pi$) 
is depicted. }
\label{fig:Fig-FS}
\end{center}
\end{figure}

\newpage 

\begin{figure}[htbp]
\begin{center}
\includegraphics[width=11.9cm]{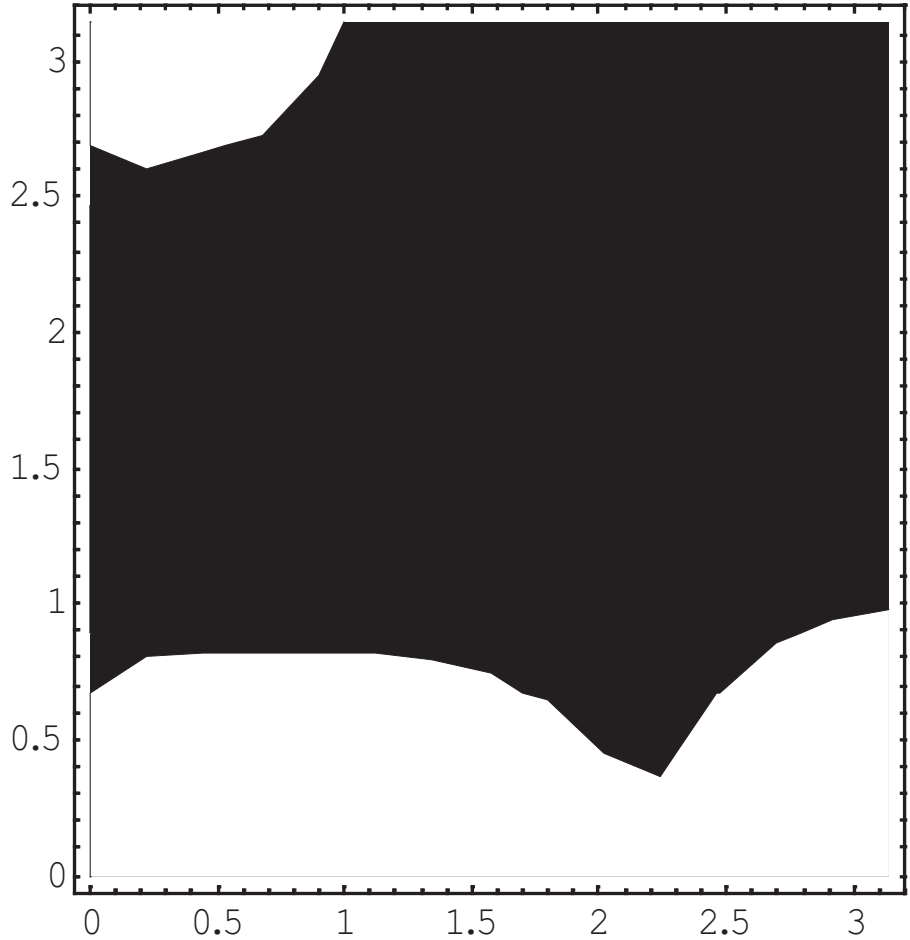}
\vskip 16mm
\caption{The factor 
$f({\bf p}) = 
 v_x ( v_x \partial v_y / \partial p_y 
     - v_y \partial v_x / \partial p_y ) $ 
in the quarter of the Brillouin zone. 
In the white region $f({\bf p}) > 0$ 
and $f({\bf p}) < 0$ in the black region. }
\label{fig:Fig-curv}
\end{center}
\end{figure}

\newpage 

\begin{figure}[htbp]
\hskip -0.8cm
\includegraphics[width=16.0cm]{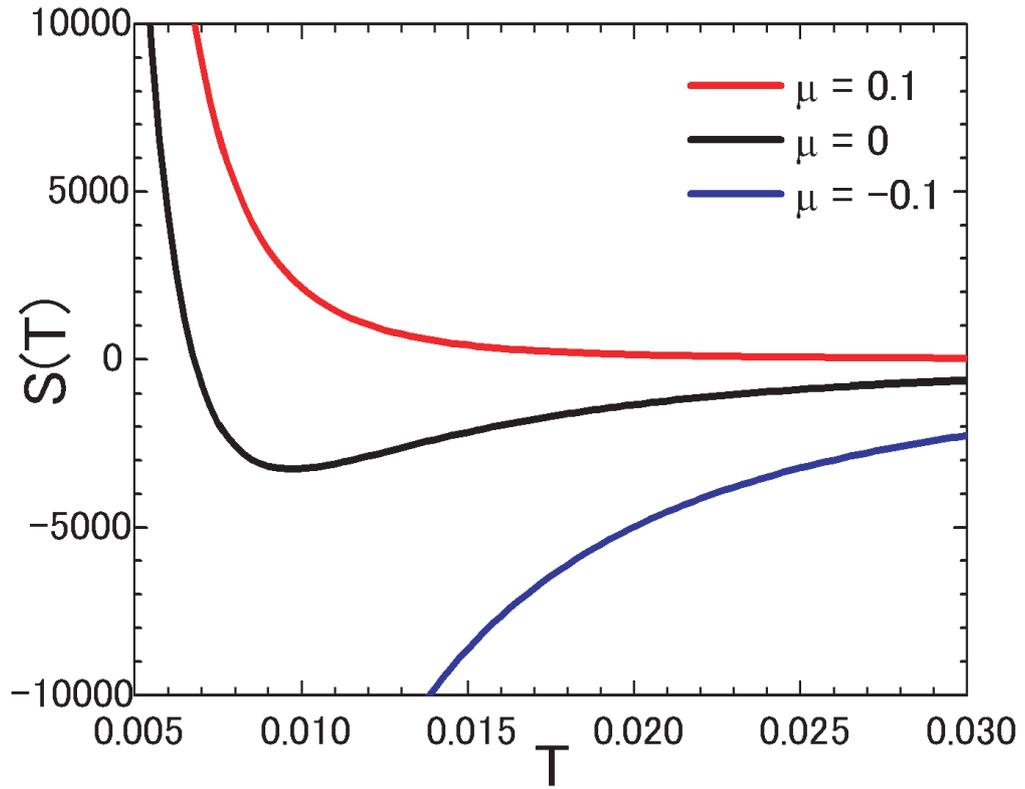}
\vskip 6mm
\caption{Temperature dependencies of $S(T)$ 
for $\mu=-0.1$, $0$, $0.1$ with $r^2 = 1.07$ 
where $S(T) \propto \sigma_{xy}$. 
The temperature is represented in eV. 
The summation over ${\bf p}$ is carried out 
using 3000$\times$3000 mesh in the quarter of the Brillouin zone. }
\label{fig:Fig-sigma}
\end{figure}

\newpage 

\begin{figure}[htbp]
\hskip -0.8cm
\includegraphics[width=16.0cm]{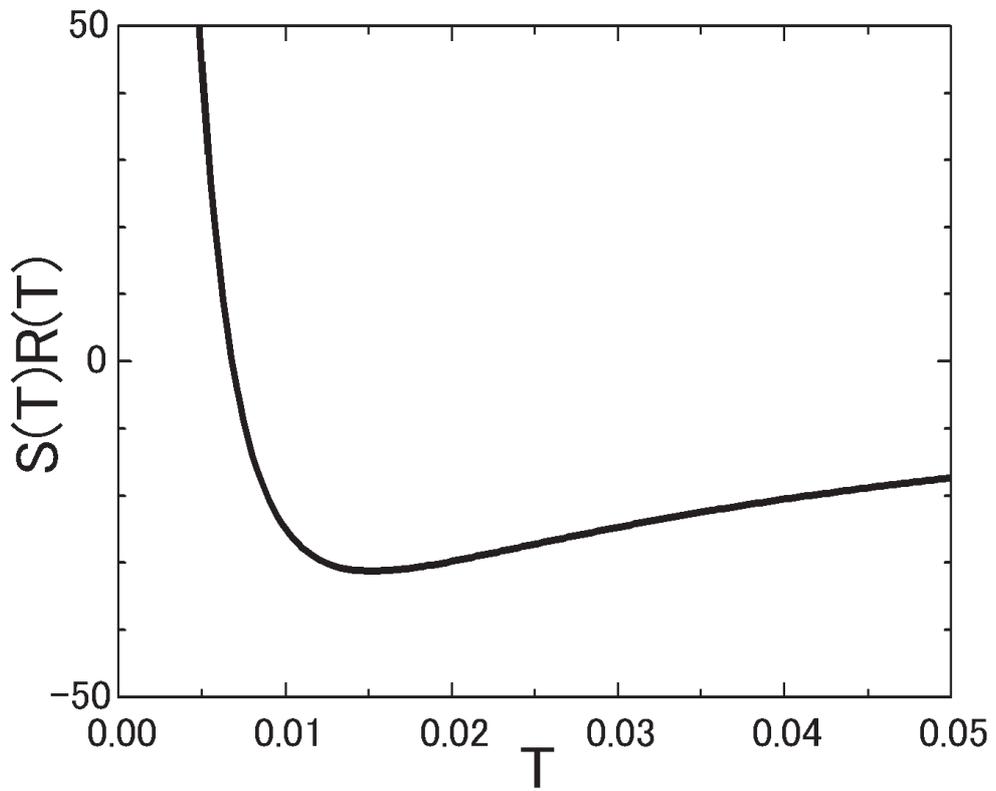}
\vskip 6mm
\caption{The temperature dependence of $S(T)R(T)$ 
for $\mu=0$ with $r^2 = 1.07$ 
where $S(T)R(T) \propto \sigma_{xy}/\sigma_{xx}$. }
\label{fig:Fig-TH}
\end{figure}

\newpage 

\begin{figure}[htbp]
\hskip -0.8cm
\includegraphics[width=16.0cm]{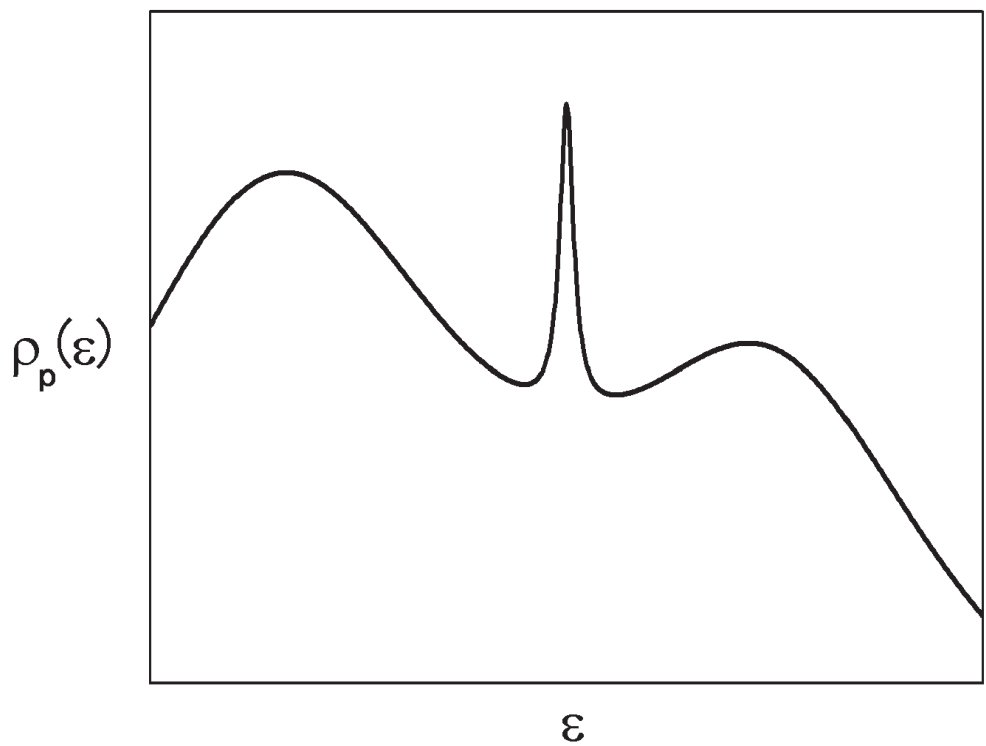}
\caption{
Energy dependence of spectral function. 
}
\label{fig:spec}
\end{figure}

\end{document}